\documentstyle[amssymb,multicol,aps,psfig,epsfig]{revtex}
\newcommand{\be}{\begin{equation}}
\newcommand{\ee}{\end{equation}}
\newcommand{\br}{\begin{eqnarray}}
\newcommand{\er}{\end{eqnarray}}
\newcommand{\nn}{\nonumber}
\newcommand{\bd}{\begin{displaymath}}
\newcommand{\ed}{\end{displaymath}}
\newcommand{\ovl}{\overline}

\newcommand{\bfig}{\begin{figure}}
\newcommand{\efig}{\end{figure}}
\def\del{\partial}

\def\alf{\alpha}

\def\lb#1{\label{#1}}
\def\3cdots{\cdot \cdot \cdot}
\def\rf#1{\ref{#1}}

\def\g{\gamma}
\def\G{\Gamma}

\def\L{\Lambda}

\def\om0{\omega _0}
\def\Om0{\Omega _0}

\def\rg{\rangle}

\def\lg{\langle}
\def\text#1{{\rm{#1}}}

\def\del{\partial}
\def\->{\rightarrow}
\def\=>{\Rightarrow}
\def\-->{\longrightarrow}
\def\==>{\Longrightarrow}

\def\rpar{\right)}
\def\lpar{\left(}
\def\lbk{\left[}
\def\rbk{\right]}
\def\lbr{\left\{}
\def\rbr{\right\}}
\def\dag{\dagger}

\def\til{\tilde}

\def\pr{^\prime}
\def\pr2{^{\prime\prime}}
\def\rf#1{(\ref{#1})}

\def\bfig{\begin{figure}}
\def\efig{\end{figure}}
\def\Tr{{\rm Tr}}

\begin{document}
\title{A consistent quantum model for continuous photodetection processes}
\author{M C de Oliveira\thanks{E-mail: marcos@df.ufscar.br},
S S Mizrahi\thanks{E-mail: salomon@df.ufscar.br}
and V V Dodonov\thanks{E-mail: vdodonov@df.ufscar.br}
\thanks{On leave from Moscow Institute of Physics and Technology
and Lebedev Physical Institute, Moscow, Russia}}
\address{Departamento de F\'{\i}sica, CCET, Universidade Federal de S\~{a}o
Carlos,\\ Via Washington Luiz km 235, 13565-905, S\~ao Carlos, SP,
Brazil.}
\date{\today}
\maketitle
\begin{abstract}
We are modifying some aspects of the continuous photodetection
theory, proposed by Srinivas and Davies [{\em Optica Acta} 1981
{\bf 28} 981], which describes the non-unitary evolution of a
quantum field state subjected to a continuous photocount
measurement. In order to remedy inconsistencies that appear in
their approach, we redefine the `annihilation' and `creation'
operators that enter in the photocount superoperators. We show
that this new approach not only still satisfies all the
requirements for a consistent photocount theory according to
Srinivas and Davies precepts, but also avoids some weird result
appearing when previous definitions are used.

pacs: {03.67.-a, 03.65.Bz, 42.50.-p}

\end{abstract}
\begin{multicols}{2}
%
\section{Introduction}
The subject of quantum measurements is as old as the very foundation of
quantum mechanics. For a long time the scheme proposed by von Neumann
\cite{Neumann} has been prevalent. According to this scheme,
there are two kinds of evolution of quantum states:
a unitary evolution obeying the rules of quantum dynamics in the absence
of measurements,
and an `extra-quantum' dynamics resulting from a measurement,
when the state vector suffers a sudden, instantaneous
and irreversible transformation (or reduction, or collapse)
to one of possible eigenstates compatible with the observable measured.

When one considers, for instance, a photocount process, the
radiation impinges a photomultiplier tube and each read burst of
electrons (current) is viewed as a manifestation of a single
photon. A sequence of bursts is associated to photocounts. A classical
theory describing this process was proposed by Mandel \cite{mandel},
however it resulted, under certain particular circumstances, in negative
probabilities of counts.
Thence refined quantum photocount theories were developed by
Mandel, Wolf and Sudarshan \cite{mandel63,mandel2}, Glauber \cite{glauber},
Kelley and Kleiner \cite{kk}, Mollow \cite{mollow}, Scully and Lamb
\cite{scully}, and others (see, e.g., the review \cite{PL-rev} for more
references).

However these theories relied on the assumption of instantaneous
measurement, although in reality photons are counted sequentially, one by
one. In the 1980's Srinivas and Davies \cite{davies} developed a theory
describing actual photocounting events,
which was adopted for estimating the
outcome in several problems, such as quantum non-demolition measurements
\cite{milburn1,milburn2}, determination of field states under
continuous photodetection process \cite{ueda}, quantum theory of
field-quadrature measurements \cite{wiseman}, conditional generation
of special states \cite{Perin96}, and for the control of the amount of
entanglement between two fields \cite{marcos}.

Srinivas and Davies (SD) theory considers photodetection as a
continuous measurement, with no reference to a `meter state'. 
The main quantities to be calculated in this theory are
probability distributions for counts (or no counts).
It is based on the assumption that in an infinitesimal time interval $\tau$
only two processes may occur: either one-count, characterized by a
superoperator $J$ acting on the field density operator $\rho$, or no-count,
characterized by another superoperator, $S_\tau$.
Superoperator $J$ in the SD theory has the form
\be%
J\rho=\gamma a\rho a^\dagger \lb{J}
\ee%
where $\gamma$ is the detector efficiency and $a$ ($a^\dagger$) is
the field `annihilation' (`creation') operator ($[a,a^\dagger]=1$).
Such a choice is based on the assumption that operator $a$ subtracts
one photon from the field. It is assumed \cite{ueda} that just after
one count (conceived as {\em subtraction} of a single photon from the
field) the system state is given by
\be%
\rho(t^+)=\frac{J\rho(t)}{\Tr[J\rho(t)]},
\label{subst}
\ee%
where $t^+$ stands for $t$ plus as infinitesimal time after.
(From another point of view, the states described by means of the
statistical operators of the form (\ref{subst}) were considered in
Refs. \cite{Dakna,LuH,Wang00} under the name `photon-subtracted states'.)

However, one can easily check that the mean number of photons in the
state (\ref{subst}) (i.e., after counting one photon) equals
\be%
{\ovl n}(t^+)={\ovl n}(t)+\lbk \frac{\ovl{\Delta n^2}(t)-\ovl{ n}(t)}
{{\ovl n}(t)} \rbk = {\ovl n}(t) + q,
\label{nt+}
\ee%
where $\ovl{\Delta n^2}=\ovl{n^2}-{\ovl{ n}}^2$ and $q$ is Mandel's
$q$ parameter \cite{Man-Q} characterizing the type of photon
statistics in the initial state of field: for $q<0$ ($q>0$) the field
statistics is said to be sub-Poissonian (super-Poissonian) and for
$q=0$ the statistics is Poissonian. However, equation (\ref{nt+}) clearly
shows that the mean number of photons {\em may increase} after one
count, if the field statistics is super-Poissonian. Otherwise, the
mean number will remain the same if the statistics is Poissonian, and
will decrease only if the statistics is sub-Poissonian. For example,
for a field represented by a Fock state $|m\rg$, one gets ${\ovl
n}(t^+)= m-1$ , exactly one photon less, whereas for any other field
state this is not true. This point received special attention in
\cite{mizdod}. Thus, we see that in general the common choice for $J$
does not really correspond one count to one less photon in the field.
Similar observations were made, e.g., in \cite{ueda,LuH,Bal79,Lee},
however, without attempts to modify the theory of photocounting
processes.

Besides, Srinivas and Davies themselves \cite{davies}
perceived that the superoperator $J$ does
not satisfy assumption (V) of their theory, namely the boundedness
property Tr$\lpar J\rho \rpar < \infty$. In fact, $J$ is an unbounded
linear transformation and consequently the counting rate is unbounded,
thus not defined for all possible states. This fact leads to an
ill-defined coincidence probability density. We shall return to this
point in the next section.

In this paper we propose some modifications in the SD photocount
theory in order to satisfy {\em all\/} the assumptions proposed by
its authors. Our motivation finds ground on the recent discussion
about the role of the `annihilation' operator $a$ in quantum optics
\cite{mizdod}, since state $a| \psi \rg $ is not always a state whose
mean number of photons is less than in $| \psi \rg $. Depending on
the field statistics, state $a| \psi \rg $ may show much higher mean
number of photons than state $| \psi \rg $. In \cite{mizdod} the
authors suggested that instead of $a$ and $a^{\dagger}$ the {\it
exponential phase operators} $E_-$ and $E_+$ should be considered as
real `annihilation' and `creation' operators in the photocounting
theory. The introduction of these operators in the continuous
photocounting theory, besides eliminating inconsistencies in the SD
proposal, leads to new interesting  results related to the counting
statistics.

This paper is organized as follows. In section 2 we briefly revise the
main aspects of quantum photodetection theories. In section 3 we
present our model based on exponential phase operators. In section 4 we
consider the field state evolution under continuous monitoring, but
when no information about the number of counted photons is read out,
or the `pre-selection' state evolution. Also, in this section we
solve the master equation generated by the exponential phase
operators instead of the annihilation/creation ones (at zero
temperature) and consider several important special cases. In section 5
we discuss a physical meaning of the results obtained, pointing at
the principal differences in the predictions of the SD theory and our
model, which could be verified experimentally.
%
\section{Fundamentals of conventional quantum photodetection theories}
%
The first quantum photodetection theory, developed independently by
Mandel {\em et al\/} \cite{mandel63,mandel2}, Glauber \cite{glauber},
and Kelley and Kleiner \cite{kk} as a simple extension of the
classical theory \cite{mandel}, gave the following probability of $k$
counts:
\br%
P(k,t) &=&{ \rm Tr}\left\{\rho : \frac{1}{k!}\left[\gamma t
I(t)\right]^k e^{-\gamma t I(t)}:\right\} \nonumber \\
&=& \sum_{n=k}^{\infty} {k
\choose n}\lpar 1 -\g t\rpar ^{n-k} \lpar \g t \rpar ^{k} p_n \,.
\lb{contagem1}
\er%
Here $\rho$ is the statistical operator of the field (we consider
a simplified model of a one-mode field),
$\gamma$ is the detector efficiency,
$I(t)$ is the average field intensity, $::$ stands for operator normal
ordering and $ p_n = \lg n| \rho |n \rg$ is the probability to have $n$
photons in the given field mode. However formula (\ref{contagem1})
becomes obviously meaningless if $\gamma t >1$, when it can result in
negative probabilities or unlimited
mean number of counted photons as $t \to \infty$.
These troubles were removed in studies
\cite{mollow,scully,davies}, whose authors, using different approaches,
arrived at the same result, which consists, from the formal point
of view, in the substitution $\g t \to 1- \exp(-\g t) $ in
equation \rf{contagem1}:
\be%
P^{SD}(k,t)=
\sum_{n=k}^\infty {n\choose{k}} (1-e^{-\gamma t})^k (e^{-\gamma
t})^{n-k}\langle n|\rho|n\rangle.
\label{PSDk}
\ee%

The SD photocount theory does not refer to a specific detector state.
It was built by considering two kinds of
events represented by superoperators acting continuously on the
field state. The first one, represented by $J$, is a single instantaneous
count event, while the other, represented by $S_{t}$,
is a no-count event for a time interval $t$. Thus, the operation (a
superoperator)
\br%
N_t(k)&=&\int_0^tdt_k\int_0^{t_k}
dt_{k-1}\cdot\cdot\cdot\nonumber\\
&&\times\int_0^{t_2}dt_1
S_{t-t_k}J S_{t_k-t_{k-1}}\cdot\cdot\cdot J S_{t_1}
\label{Ntk}
\er%
stands for the count 
of $k$ photons from the field.
The operator (\ref{Ntk}) projects continuously the initial field state
$\rho$, and $P(k,t)=\Tr[N_t(k)\rho]$ is the
probability of counting exactly $k$ photons in a time interval $t$.

The assumptions or properties of the SD theory are as follows.

\noindent (I) $\rho\rightarrow N_t(k)\rho$ is a linear positive map
on the space ${\cal T(H)}$ of trace class operators on the Hilbert
space ${\cal H}$, such that for any positively definite $\rho$
satisfying $\mbox{Tr}\,\rho=1$, one has
\be%
0\le \Tr[N_t(k)\rho]\le 1.
\ee%
\noindent (II) The sum over all possible counted photons satisfies
the normalization condition Tr$[T_t\rho]=1$ where
%
\be%
 T_t=\sum_{k=0}^\infty N_t(k) = S_t+ \int_0^t T_{t-t'}JS_{t'} dt'.
\ee%
\noindent (III) Semigroup associative property
\be%
N_{t_1+t_2}(k)=\sum_{k_1+k_2=k} N_{t_2}(k_2) N_{t_1}(k_1).
\ee%
\noindent (IV) Identity
\be%
\lim_{t\rightarrow 0} N_t(0)\rho=\rho.
\ee%
\noindent (V) Assumption of bounded interaction rate: there exists a
number $K<\infty$ such that
\be%
\sum_{k=1}^\infty \Tr[N_t(k)\rho]<K t, \mbox{ for } t>0.
\ee%
\noindent (VI) Assumption of ideality: the operation \be S_t=N_t(0)
\ee transforms pure states into pure states, noting that $ S_t \rho$
is a pure state if $\rho$ is a pure state.

As a matter of fact the process of photocounting by a macroscopic
detector is the resultant of many microscopic fundamental
interactions between the electromagnetic field and atoms
composing the detector. But how these fundamental interactions can
be specified by a reasonable model for the count (or no-count)
jump super-operator in a phenomenological macroscopic
photodetection theory?
It is worth citing the corresponding words from the Srinivas and Davies's
paper \cite{davies}:
``{\em As a simple model for the measurement performed by the
photodetector, where an $n$-photon state will be converted to an
$(n-1)$-photon state whenever a photon is detected, it is reasonable
to set\/}''
the single photon counting superoperator
$J$ in the form  \rf{J}.
We see that this operator was not derived from fundamental processes,
but set {\em ad hoc} due to its apparent simplicity and reasonability.

As soon as $J$ is chosen, the superoperator for the
state evolution between consecutive counts can be derived as
\be
S_t \rho=e^{Yt}\rho e^{Y^\dagger t}, \qquad
Y=-iH_0- \g a^\dagger a /2,
\lb{StSD}
\ee
once it conserves the probability in a regular point process:
\be%
\label{11} \Tr[J\rho+Y\rho+\rho Y^\dagger]=0 .
\ee%

Note, however, that the superoperators $J$ and $S_t$ defined above
{\em do not satisfy} condition (V), since
\be%
{\Tr}(J\rho)=\gamma {\Tr}(\rho a^\dagger a),
\ee%
or more generally,
\be%
{\Tr}[J^k\rho]=\gamma^k {\Tr}[\rho :n^k:]
\ee%
is unbounded, so, not defined for all states. The violation of
assumption (V) prevents a consistent definition for all states of
some important functions. For example, although one can define the
elementary probability density of counting $k$ photons in a time
interval $t$, attempts to calculate the coincidence probability
density
\be%
h(t_1,...,t_k)={\Tr}(T_{t-t_k}J...JT_{t_1}\rho)
\ee%
of counts observed at each of the times $t_1,...,t_k$ together with
other possible counts in between, when the detector is making
measurements for a time interval $t$, result in serious problems in
the SD theory.

One of the goals of our paper is to show how one can avoid the
violation of assumption (V) using other operators instead of $a$ and
$a^\dagger$.
%
\section{Quantum counting processes with exponential phase operators}
%
Our idea is to use, instead of operators $a$ and $a^{\dagger}$, the
so-called {\em exponential phase operators}
\br%
E_-&\equiv& (a^\dagger a+1)^{-1/2} a,
\label{E-} \\
E_+&\equiv& a^\dagger(a^\dagger a+1)^{-1/2},
\label{E+}
\er%
introduced, as a matter of fact, by F London at the dawn of quantum
mechanics \cite{London}, although their systematic use began only
after the paper by Susskind and Glogower \cite{susg} (for the history
see \cite{Niethis}). The commutator of $E_-$ and $E_+$ is the vacuum
state projector,
\be%
[E_-,E_+]= |0\rg\lg 0| \equiv \L_0.
\ee%
The normal ordered product of these operators is a complementary projector
\be%
E_+E_-=1-\L_0 = \L =\L^{\dagger}, \qquad \L ^2 =\L.
\label{defLam}
\ee%
Applying $E_-$ and $E_+$ on the number states one gets
\be%
E_- |n\rangle = |n-1\rangle\, , \quad E_+ |n\rangle = |n+1\rangle
\ee%
with $E_- |0\rangle = \L |0\rg = 0$, and $[\L,n]=0$
(where $n=a^{\dagger}a$ is the number operator). A useful property is
\be%
e^{\alf \L} = \L_0 + e^\alf \L.
\label{propLam}
\ee%
For other properties and generalizations see, e.g.,
\cite{CarNiet,Paul,Imoto85,BP86,BeEng91,Loudon,LukPer,Ban95,Lynch95,Royer96}.
Finally, it is worth to recall that $E_-$ has as eigenstate the
`coherent phase state'
\be%
|\psi\rangle=\sqrt{1-|z|^2}\sum_{n=0}^\infty z^n |n\rangle, \quad |z|<1
\ee%
as introduced in \cite{Ler} and studied in
\cite{Ifan,ShapShep,ChaKa91,Vourd92,Hall93,Sudar93,BrifBen94,%
DoMi,VBM96,Wun01} (see also \cite{rev-ncs}).
%
\subsection{One-count event}
%
We redefine the one-count operator equation \rf{J} as
\be%
J\rho = E_-\rho E_+ \,.
\label{J-E}
\ee%
Now $J$ is bounded operator and the system state immediately after
the 1-count process in the time interval $[0,t)$ is transformed into
\be%
\label{1} \til \rho(t^+)=\frac{J\rho(t)}{\Tr [J\rho(t)]}
=\frac{J\rho(t)}{1-p_0}\,,
\ee%
where $p_0 \equiv\langle 0|\rho(t)|0\rangle$ is the probability for
the vacuum state. (Note that pure states $\hat{E}_+^m|\psi\rangle$
were considered in another context in \cite{Moya99}. Mixed {\em
shifted thermal states}
$\hat\rho_{th}^{(shift)}=\hat{E}_+^{m}\hat\rho_{th}\hat{E}_-^{m}$
were studied in \cite{Lee97}, whereas methods of generating such
states in a micromaser were discussed in \cite{SMW96}.)

The mean number of photons in the state $\til \rho(t)$ (\ref{1}) is
\be%
{\til n}(t^+)=\frac{{\ovl n}(t)}{1-p_0} - 1,
\ee%
so, whenever a state $\rho$ has none, or very small, contribution
from the vacuum state, the counting operation extracts exactly one
photon from the system, independently of the field statistics. For
example, for the {\em number state} $\rho = |m\rg \lg m |$ ($m\neq
0$), ${\til n}=m-1$, and for the {\em coherent state} $\rho =
|\alpha\rg \lg \alf |$ ($\alpha\neq 0$), ${\til n}={\ovl
n}/\left(1-e^{-\ovl n}\right) -1$, with ${\ovl n}=|\alpha|^2$.

On the other hand, for the {\em thermal state}
\be%
\label{thermal}
\rho = \frac {1}{1+\ovl n}\sum_{n=0}^\infty\left(\frac{
\ovl  n}{1+ \ovl n}\right)^n |n\rangle\langle n|
\ee%
we obtain $\til{n}= \ovl{n}$, i.e., the mean number of photons is not
changed, as expected. This is a correct description of a thermal
system since taking out a single photon from a reservoir should not
change its average number. Note that using the SD definition, $J\rho
= a \rho a^{\dagger}$, one obtains the weird result $\til n = 2
\ovl{n}$. So one perceives that using $a$ and $a^{\dagger}$ for
constructing a continuous photocount measurement leads to some
inconsistent results.
%
\subsection{No-count event}
%
The time evolution between sequential counts is represented by $
S_t\equiv N_t(0)$, a superoperator defined in terms of ordinary
Hilbert space operators
\be%
S_t \rho =e^{Yt} \rho e^{Y^\dagger t},
\ee%
where $B_t = e^{Yt}$ is a semigroup element given in terms of the
generator $Y$. The deduction of $S_t$ is conditioned to the relation
\be%
\label{2} \Tr[J\rho]= \Tr[\rho R],
\ee%
where $R$ is the rate operator, related to $Y$ by $\Tr(\rho R) = \Tr
(Y\rho+\rho Y^\dagger)$, which substituted in (\ref{2}) gives equation
(\ref{11}). The theory requires that in the absence of counts the
system has a unitary evolution, whose dynamics is governed by the
free-field Hamiltonian $H=\hbar\omega a^\dagger a$. Thus, the
convenient choice satisfying (\ref{11}) is
\be%
Y= -iH - R/2 = -iH- \frac\gamma 2 E_+E_-\,.
\ee%

Taking into account equations (\ref{defLam}) and (\ref{propLam}), as well
as the commutativity of operators $H$ and $\Lambda$, one can easily
calculate the result of action of nonunitary operator $e^{Y\tau}$ on
a pure state $| \psi \rg$ (here $\tau$ is an interval of time between
counts)
\be%
|\psi_S (\tau)\rg = e^{Y\tau}|\psi \rg =  \lg
0| \psi \rg |0 \rg + e^{-\g \tau/2} \L |\psi_H(\tau) \rg,
\ee%
where $|\psi_H(\tau) \rg =\exp(-i H\tau)$ is the freely evolved state
vector (in the absence of measurements). Thus the probability of {\em
no-count\/} event equals
\be%
{\cal P}_0(\tau)= \Vert\,|\psi_S (\tau)\rg\,\Vert= |\lg0|\psi \rg |^2
+ e^{-\g \tau} \lg \psi | \L | \psi \rg . \label{prob0pur}
\ee%
For a mixed state, the same probability is given by $\Tr[S_\tau \rho]$,
and simple calculations result in the formula
\be%
{\cal P}_0(\tau)=e^{-\gamma\tau}+p_0(1-e^{-\gamma\tau}),
\label{prob0}
\ee%
which, of course, coincides with (\ref{prob0pur}) in the case of pure
state. Note that $\lim_{\tau \to \infty}{\cal P}_0(\tau) = p_0 $,
which means that the probability of no counts registered during an
infinite time interval is equal to the probability of finding the
vacuum state in the measured state $\rho$. Formula (\ref{prob0})
should be compared with analogous formula of the SD theory based on
the operator $Y$ of the form (\ref{StSD})
\be
{\cal P}_0(\tau)=\sum_{n=0}^{\infty} p_n e^{-n\gamma\tau}.
\label{prob0SD}
\ee
Although equations (\ref{prob0}) and (\ref{prob0SD}) give the same
limits for $\tau \to \infty$, the intermediate time dependencies are
different.
%
\subsection{Continuous counting}
%
The continuous counting of $k$-photons from a field in a time
interval $t$ is represented by a linear operator $N_t(k)$ acting on
the system state during a time interval $[0,t)$,
\be%
\til \rho^{(k)}(t)=\frac{N_t(k)\rho (0)}{{\Tr}\lbk N_t(k)\rho(0)
\rbk },
\ee%
where $\rho (0)$, or simply $\rho$, is the state of field prior to the
counting process and $P(k,t)= \Tr (N_t(k)\rho)$ is the probability of
counting $k$ photons in $t$. The linear operator $N_t(k)$ can be written
in terms of the operators $S_t$ and $J$ as
\br%
\label{3} N_t(k)&=&\int_0^tdt_k \int_0^{t_k} dt_{k-1}\cdots
\int_0^{t_2} dt_1  \nonumber \\ && \times
S_{t-t_k}JS_{t_k-t_{k-1}}\cdots JS_{t_1}\, .
\er%
Noticing however that
\be%
JS_t \rho = e^{- \g t} {\cal U}_t (J \rho),
\ee%
one gets
\be%
S_{t-t_k}JS_{t_k - t_{k-1}} \cdot \cdot \cdot J S_{t_1} = e^{- \g
t_k}{\cal U}_{t_k} S_{t-t_k} J^k,
\label{SJS}
\ee%
where
\be%
{\cal U}_t \rho = e^{-iH t} \rho e^{iH t}.
\label{Utr}
\ee%

It is convenient to introduce short notation for two partial sums
of probabilities:
\be%
{\cal A}_k=\sum_{n=0}^{k} p_n , \quad {\cal
Z}_{k+1}=\sum_{n=k+1}^{\infty} p_n \equiv 1- {\cal A}_k\,.
\label{defAZ}
\ee%
Using (\ref{SJS}) we can calculate the elementary probability
distribution (EPD) of counts at the instants $t_1,t_2,\ldots,t_k$,
if the total measurement time is $t$,
\br%
{\cal P}(t_1, t_2, \cdots, t_k; t)  &\equiv&  \Tr \lbk S_{t-t_k} J
S_{t_k-t_{k-1}}, \cdots , J S_{t_1} \rho \rbk \nonumber \\ &=& \g ^k
\lpar e^{- \g t_k}  p_k + e^{- \g t }{\cal Z}_{k+1} \rpar, \lb{epd}
\er%
which should be compared to the EPD in SD theory,
\br%
{\cal P}_{SD}(t_1, t_2, \cdots, t_m; t)  &=&
\g ^m m! e^{-\g (t_1 + t_2 + \cdots + t_m -mt) }
\nonumber \\ && \times
\sum_{n=m}^{\infty}  {n \choose m} e^{- \g nt} p_n\,.
\lb{epdsd}
\er%
In particular, for $t=\infty$ we obtain
\be%
{\cal P}(t_1, t_2, \cdots, t_k; \infty) = \g ^k  e^{- \g t_k}  p_k\,,
\label{epd-E}
\ee%
whereas the SD theory yields essentially different result for $k\ge 2$
\be%
{\cal P}_{SD}(t_1,  \cdots, t_m; \infty)  =
\g ^m m! e^{-\g (t_1 + t_2 + \cdots + t_m ) } p_m\,.
\label{epd-a}
\ee%
Only for the one-photon event, $k=1$, both models predict the same
exponential probability distribution
\[
{\cal P}(t_1)= \g e^{- \g t_1}  p_1\equiv \tilde{\cal P}(t_1) p_1.
\]
 We see that \rf{epd} corresponds to a Markovian process, in the sense that
\[
{\cal P}(t_1, t_2, \cdots, t_k; \infty)/ p_k =
\tilde{\cal P}(t_1)\tilde{\cal P}(t_2-t_1)\cdots
\tilde{\cal P}(t_k-t_{k-1}),
\]
i.e., the EPD depends only on the last count, at time $t_k$.
On the contrary, in the SD theory the EPD depends on all times
at which counts occur, moreover, each new count enters with increasing
weight:
\[
{\cal P}_{SD}(t_1, t_2, \cdots, t_m; \infty)/ p_k  =
\tilde{\cal P}(t_1)\cdot2\tilde{\cal P}(t_2)\cdot\,\ldots\,\cdot
m\tilde{\cal P}(t_m).
\]
Nonetheless, both distributions, (\ref{epd-E}) and (\ref{epd-a}),
have the same normalization
\[
\int_0^{\infty}dt_k \int_0^{t_k}dt_{k-1}
\cdots \int_0^{t_{2}}dt_{1}
{\cal P}(t_1, t_2, \cdots, t_k; \infty) = p_k\,.
\]

We can write equation (\ref{3}) as ($k=1,2,\ldots$)
\br%
N_t(k) \rho &=& {\cal U}_t  \int _ 0 ^t dt' e^{- \g t'}
\frac{\lpar t'\rpar ^{k-1}}{(k-1)!}
\exp\left[-\,\frac{\g}{2} (t-t') \L\right]
\nonumber \\ && \times
\lpar J^k \rho\rpar
\exp\left[-\,\frac{\g}{2} (t-t') \L\right] ,
 \lb{ntk}
\er%
\noindent where  $t' \equiv t_k $ and the probability of occurrence
of $k$ counts in a time interval $t$ is given by
\br%
P(k,t)&=& p_k  \lbk 1- e^{-\gamma t} \sum_{j=0}^{k} \frac {(\gamma
t)^j}{j!} \rbk \nonumber \\ &&
+ \frac{ e^{-\gamma t} (\gamma t)^k}{k!}
\sum_{n=k}^\infty p_n , \label{Pkk}
\er%
which verifies the normalization condition
\be%
\sum_{k=0}^\infty P(k,t)=1 .
\ee%
The limiting value of equation (\ref{Pkk})
\be%
\lim_{t \to\infty} P(k,t)= \langle k|\rho|k\rangle = p_k \label{limPk}
\ee%
means that, asymptotically, the counting statistics coincides with
the photon statistics (as it should), since we do not consider
here the possibility of photons lost to the surroundings or the
failure to count any photon exiting the cavity. If a photon leaves
the cavity it is detected and counted, for sure.

The moments of distribution \rf{Pkk} are given by
\br%
&&{\ovl {k^l}_t} \equiv \sum_{k=0}^{\infty} P(k,t) k^l
= \lg k^l \rg
\nonumber \\ &&
 - e^{-\gamma t} \lbr \sum_{j=0}^\infty
\frac{(\gamma t)^j}{j!}\sum_{k=j}^\infty k^l p_j 
- 
\sum_{k=0}^\infty \frac{k^l (\gamma t)^{k}}{k!} \sum_{j=0}^k p_j \rbr ,
\er%
where $\lg k^l \rg \equiv \sum_{k=0}^{\infty} k^l p_k $.
In particular, for the mean number of photons  one verifies that
$ \lim_{t \to \infty} {\ovl k_t} = \lg k \rg $.
%
\subsection{Examples of field states}
%
Now let us consider specific field states, comparing the probabilities
of $k$-counts resulting from the Srinivas and Davies formula
(\ref{PSDk}) and from our formula (\ref{Pkk}).
For a field initially prepared in a number
state $|m\rangle$, $m>k$, the probabilities are
\br%
P^{SD}(k,t)&=&{m\choose{k}}(1-e^{-\gamma t})^k(e^{-\gamma
t})^{m-k},\label{sd}\\ P(k,t)&=&e^{-\gamma t}\frac{(\gamma
t)^k}{k!}\label{mvs}.
\er%

Different probabilities of a $k$-count reveal different physical
schemes of counting. In SD's theory, $ P^{SD}(k,t)$ is a binomial
distribution reflecting an underlying one-dimension `random walk'
process: the factor $\left(1-e^{-\gamma t} \right)^k$ is the
probability of $k$ photons leave the cavity, while $\left( e^{-\gamma
t}\right)^{m-k}$ is the probability that photons stay in the cavity,
thus being not counted. On the other hand, the distribution
(\ref{mvs}) is Poissonian. That means that each photon leaving the
cavity is counted, resulting in the same counting statistics as for
falling raindrops in a small area, roughly the drop size. The last
process is the only one consistent with the initial assumption that
every photon leaving the cavity is counted. In figure 1 we compare
both probability distributions, (\ref{sd}) and (\ref{mvs}), for a
field with $m=5$, and selected $k$-counted photon numbers. $P(k,t)$
for (\ref{mvs}) reveals a more spread shape then (\ref{sd}), a
characteristic of Poissonian processes. In that figure we included
the probability distributions for $k=m$. The SD theory gives the
expression $P^{SD}(m,t)=(1-e^{-\gamma t})^m$, while the present model
gives
$P(m,t)= \Phi_{m-1}(\gamma t)$,
where
\be%
\Phi_k(x)= 1 - e^{-x}\sum_{n=0}^k \frac{x^n}{n!} =
e^{-x}\sum_{n=k+1}^{\infty} \frac{x^n}{n!}\,. \label{defPhik}
\ee%
Notice that for $k>m$, both theories give a zero-valued probability
distribution, which is a signature of the fixed number state for the
field inside the cavity, meaning that, in any time interval $t$, it
is not possible to count more photons than those present in the field
at time $t=0$.

For the coherent state $|\alpha\rangle$ SD theory gives
\[%
P^{SD}(k,t)=\frac 1{k!}\left[|\alpha|^2\left(1\!-\!e^{-\gamma t}\right)
\right]^k
\exp\left[-|\alpha|^2\left(1\!-\!e^{-\gamma t}\right)\right],
\]%
while in our present approach we get
\[%
P(k,t)= \frac{1}{k!}\left[
(\gamma t)^k e^{-\gamma t}\Phi_{k-1}(|\alpha|^2)
+|\alpha|^{2k}e^{-|\alpha|^2}\Phi_k(\gamma t)\right].
\]%
In figure 2 we display, for comparison, both distributions for a coherent
state with average photon number $|\alpha|^2=5$, and selected counted
photon numbers.

For the thermal state (\ref{thermal}) we obtain the distributions
\[%
P^{SD}(k,t)=\frac{\left[{\ovl n}\left(1-e^{-\gamma t}\right)\right]^k}
{\left[1+ {\ovl n}\left(1-e^{-\gamma t}\right)\right]^{k+1}},
\]%
\[%
P(k,t)=\frac{e^{-\gamma t}}{k!}
\left(\frac{{\ovl n} \gamma t}{{\ovl n} +1}\right)^k
+\frac{{\ovl n}^k \Phi_k(\gamma t)}{\left(1+{\ovl n}\right)^{k+1}}\,,
\]%
which are displayed in figure 3, for ${\ovl n}=5$. For both, the
coherent and thermal states, the probability distribution for the
present model also show a more spread shape than the distribution for
the SD theory. Perhaps, this feature may be important in the
distinction of both models for photocounting, by experimental
evidence. By repeated experiments of photocounting the output of a
leaking cavity, one may reconstruct those distributions, assuming
that any other incoherent (dissipative) process is absent or
negligible for the whole process.
%
\section{`Pre-selection' state evolution}
%
Having no knowledge about the number of counted photons after a time
interval $t$, the field state being continuously monitored, but with
no readout is given by
\be%
\til{\rho} (t)=T_t\rho=\sum_{k=0}^\infty N_t(k)\rho .
\ee%
It is referred as the {\em pre-selection} state \cite{caves}.
Summing over $k$ in equation (\ref{ntk}) we obtain
\br%
&&\til{\rho} (t) =  S_t \rho + \sum _{k=1}^{\infty} N_t(k)\rho = {\cal
U}_t \lbr e^{-\g \Lambda t/2} \rho e^{-\g \Lambda t/2}
\right. \nonumber \\ && \left.
+ \int ^{t}_0
e^{-\g t'} e^{-\g \Lambda (t-t')/2} \lbk e^{Jt'} \lpar J \rho \rpar
\rbk e^{-\g \Lambda (t-t')/2} dt'  \rbr.
\er%
The probability of having $n \neq 0$ photons in the cavity after
continuous measurement for time $t$ is equal to
\be%
\til{p}_n (t) = \lg n| \til{\rho} (t) | n \rg = e^{-\g t}\sum_{l=0}^
{\infty}\frac{(\g t)^l}{l!}p_{n+l}\,, \lb{ptil}
\ee%
where $p_{n+l}$ is the probability at $t=0$. The probability to have
no photons, in the cavity, at time $t$ is obviously
\be%
\til{p}_0 (t) = 1 -\sum_{n=1}^{\infty}\til{p}_n(t) = e^{-\g
t}\sum_{l=0}^{\infty} \frac{(\g t)^l}{l!}{\cal A}_{l},
\ee%
where ${\cal A}_{l}$ is defined in (\ref{defAZ}).
So, if one waits very long time the cavity will eventually end in the
vacuum state, all photons being absorbed by the detector and counted.
%
\subsection{The master equation}
%
Another way to obtain formula \rf{ptil} is to use the phenomenological
master equation of the Lindblad form (hereafter we suppress the tilde over
the operator $\til\rho$)
\be%
\frac{\del\rho(t)}{\del t}=\frac\gamma
2[2E_-\rho(t)E_+-E_+E_-\rho(t)-\rho(t)E_+E_-].
\lb{masterE}
\ee%
It should be compared with the `standard master equation' \cite{CoTa,Carm}
for the
amplitude damping model used in the SD theory, derivable from the
interaction of a single electromagnetic (EM) mode (or 1-D
harmonic oscillator) with an environment made of many harmonic
oscillators at T$=0$K (see, e.g., \cite{Calsa01} for the most
recent applications)
\be%
\frac{\del\rho(t)}{\del t}=\frac\gamma 2
\left[2a\rho(t)a^{\dagger} -
a^{\dagger}a\rho(t)-\rho(t)a^{\dagger}a\right]. \lb{master}
\ee%
If the EM mode is within a dissipative cavity, equation \rf{master} can
be viewed as describing the field state of an uncontrollable
``photon-leaking" process to the environment, although it could be
not necessarily one-by-one, since $J= \g a\rho a^\dag$ does not
produce a minus-one-photon state. In contradistinction, if there
is no vacuum, for sure, inside the cavity, equation \rf{masterE}
describes the process of {\em subtracting photons from the cavity,
sequentially and one-by-one}.

Equation (\ref{master}) results in the equation for the mean
photon number
\be%
\label{number} - {d \langle a^\dagger a\rangle ^{(a)} }/{d t} =
\gamma \langle a^\dagger a\rangle =\g \sum_{n=1}^{\infty}n p_n,
\label{rate-a}
\ee%
whose solution
\be%
\label{ampd} \langle a^\dagger a\rangle ^{(a)} _t = \langle
a^\dagger a\rangle ^{(a)} _0 e^{-\g t} \lb{nmedioa}
\ee%
shows that the rate of decrease of the mean number of photons is
proportional to the mean number of photons present inside the
cavity (or in the beam, whether beams are considered) resulting in
an exponential decay as time goes on.

Equation (\ref{masterE}) leads to a quite different differential
equation for the mean photon number
\be%
\label{number2} - {\del \langle a^\dagger a\rangle ^{(E)} }/{\del
t}= \gamma (1 - p_0)= \g \sum_{n=1}^{\infty}p_n,
\label{rate-E}
\ee%
%
according to which, the
rate of change in the mean number of photons is proportional to
the probability that there are photons in the cavity, independently
of their mean number. If initially $p_0=0$, this means that there
are photons in the cavity for sure and equation \rf{number2} becomes
\be%
\left. - {\del \langle a^\dagger a\rangle ^{(E)} }/{\del
t}\right|_{t=0}= \G \lpar \langle a^\dagger a\rangle |_{t=0 }
^{(E)} \rpar \langle a^\dagger a\rangle|_{t=0 } ^{(E)}
\ee%
with $\G \lpar \langle a^\dagger a\rangle ^{(E)} \rpar= \g/
\langle a^\dagger a\rangle ^{(E)}$ . Or, way around, at initial
times the rate does not depend on the mean photon number due to
the nonconstant coupling parameter $\G$, following the choice of
operators $E_- $ for picking out a single photon from the field.

Note that equations (\ref{rate-a}) and (\ref{rate-E}) practically
coincide in the case of low field intensities, when $p_n \ll 1$
for $n\ge 2$ (this is especially clear from the expressions
in the form of series over $p_n$). On the other hand, the
`saturation' of the decay rate in the case of high field intensities
can be also understood, if one takes into account a possibility
of large dead time of the detector: in such a case, for short
intervals of time, the detector can count (with some efficiency)
only one photon, independently of the number of photons in the
cavity or beam. This example shows that the decay rate (\ref{rate-E})
by no means can be considered as `unphysical' {\em apriori\/}.

It is immediate to see that the solution to equation \rf{number2} is
\br%
\langle a^\dagger a\rangle ^{(E)} &=& \sum_{n=1}^\infty
n\tilde{p}_n (t) = e^{-\g t}\sum_{n=1}^\infty n \sum_{l=0}^
{\infty}\frac{(\g t)^l}{l!}p_{n+l} \nn \\ &=& e^{-\g t}\lbk
\sum_{n=1}^\infty n p_n + (\g t)\sum_{n=1}^\infty n p_{n+1} +
\3cdots \rbk \\
&=& e^{-\g t}\lbk \bar n +  (\g t)\sum_{n=1}^\infty n p_{n+1} +
\3cdots \rbk ,  \lb{nmedioE}
\er%
However both equations, \rf{number} and \rf{number2} have the
vacuum state as the asymptotic stationary state. Subtraction of
\rf{nmedioa} from \rf{nmedioE} leads to
\be%
\langle a^\dagger a\rangle ^{(E)} _t - \langle a^\dagger a\rangle
^{(a)} _t = e^{-\g t}\sum_{l=1}^{\infty}\frac{(\g t)^l}{l!} \sum
_{n=1}^{\infty} n p_{n+l} >0, \lb{difmean}
\ee%
which shows that by using the nonlinear operators $E_-,E_+$, the
calculated mean photon number of the remaining photons in the
cavity is always higher than when $a, a^\dag$ operators are used.
Because, with $E_,E_+$-operators, the photons are always
subtracted sequentially, one-by-one from any field state that does
not contain the vacuum as one of its components, while by using
$a, a^{\dagger}$ operators, and depending on the field state,
there exists the possibility that the photon-leak occurs through
the escape of more than one photon at a time. Thus, mean photon
number reduction rate for $E$-operators does not follow an
exponential law, it lasts a longer time to have the cavity
reaching a given photon mean photon number than in the amplitude
damping model. We will touch this point again in the next section
for a few specific examples of field states.
%
\subsection{Solutions of the master equation}
%
The operator master equation (\ref{masterE}) results in the following
infinite set of coupled equations for the diagonal elements
$p_n\equiv \langle n|\rho| n \rangle$
of the statistical operator in the Fock basis:
\be%
\dot{p}_n= p_{n+1} -\left(1-\delta_{n0}\right) p_n,
\label{eqdotpn}
\ee%
where dot means the derivative with respect to the `slow time'
$\tau=\gamma t$. Making the Laplace transformation
\[%
\overline{p}_n(s)=\hat{\cal L}\left\{p_n(t)\right\}
= \int_0^{\infty}e^{-s\tau}p_n(t)d\tau
\]%
and remembering that
\[%
\hat{\cal L}\left\{\dot{p}_n(\tau)\right\}=
s\overline{p}_n(s) -p_n(0),
\]%
we obtain from (\ref{eqdotpn}) the following set of algebraic equations:
\br%
\overline{p}_{n+1}(s)&=& (s+1)\overline{p}_n(s) -p_n(0), \quad n\ge
1, \label{Lapeqn} \\ \overline{p}_{1}(s) &=& s\overline{p}_0(s)
-p_0(0). \label{Lapeq1}
\er%
Their consequence is the relation
\br%
\overline{p}_{k}(s)&=& s(s+1)^{k-1}\overline{p}_0(s) -(s+1)^{k-1}p_0(0)
\nonumber \\ &&
-(s+1)^{k-2}p_1(0) -\cdots - p_{k-1}(0). \label{recrel}
\er%
Suppose that only the states with $n\le k$ were excited initially.
Then $p_{k+1}(\tau)\equiv 0$ and consequently
$\overline{p}_{k+1}(s)\equiv 0$. In this case the consequence of
(\ref{recrel}) is
\br \overline{p}_{0}(s) &=& \frac{p_0(0)}{s} + \frac{p_1(0)}{s(s+1)} +
\frac{p_2(0)}{s(s+1)^2} +\cdots
\nonumber \\ &&
 + \frac{p_k(0)}{s(s+1)^k}\,.
\label{Lappos}
\er%
Taking into account the relations
\br%
\hat{\cal L}\left\{\exp(-at)\right\}&=&(a+s)^{-1}, \nn \\ \hat{\cal
L}\left\{t^m e^{-t}\right\}&=& m!(s+1)^{-m-1}, \nn
\er%
and the expansion
\[%
\frac1{s(s+1)^{k}}= \frac1{s} - \frac1{(s+1)} - \frac1{(s+1)^{2}}
- \cdots - \frac1{(s+1)^{k}}\,,
\]%
we can find the inverse Laplace transform of (\ref{Lappos}): 
\br%
p_0(\tau) &=& p_0(0) + p_1(0)\left[1\!-\!e^{-\tau}\right] + p_2(0)
\left[1\!-\!e^{-\tau}(1\!+\!\tau)\right] \nonumber \\ &&
+ \cdots
+ p_k(0) \Phi_{k-1}(\tau),
\label{solp0}
\er%
where the function $\Phi_{k}(x)$ was defined in (\ref{defPhik}).
Having arrived at the expression (\ref{solp0}), one can verify that it
does not depend on the initial auxiliary assumptions that $p_n(0)=0$
for $n>k$, but it holds for any initial distribution. Knowing
$p_0(\tau)$ one can find all other function $p_n(\tau)$ from
equations (\ref{eqdotpn}).
Finally, one arrives at the expression
which coincides exactly with (\ref{ptil}):
\be%
p_m(\tau)=e^{-\tau}\sum_{k=0}^{\infty}\frac{\tau^k}{k!}p_{m+k}(0),
\quad m\ge 1.
\label{solm}
\ee%

The reduced generating function of the diagonal matrix elements
depends on time as follows,
\be%
\tilde{G}(z;\tau)\equiv \sum_{n=1}^{\infty}z^n p_n
= e^{-t} \sum_{k=1}^{\infty}p_k(0)
\sum_{n=1}^{k}\frac{z^n \tau^{k-n}}{(k-n)!}\,.
\label{Gen1}
\ee%
Thus the mean number of photons evolves as
\be%
\overline{n}(\tau)=\left.\frac{\partial\tilde{G}}{\partial z}\right|_{z=1}
= e^{-t} \sum_{k=1}^{\infty}p_k(0)
\sum_{n=1}^{k}\frac{n \tau^{k-n}}{(k-n)!}\,.
\label{avn}
\ee%
For the initial $k$-photon Fock state $|k\rangle$ we obtain
\be%
p_m^{(Fock)}(\tau)=e^{-\tau}\frac{\tau^{k-m}}{(k-m)!}, \quad 1\le
m\le k, \label{pmFock} \ee
\be
\overline{n}^{(Fock)}(\tau)= e^{-\tau} \sum_{n=1}^{k}\frac{n
\tau^{k-n}}{(k-n)!}\,. \label{avFock}
\ee%

\subsection{Special cases}
%
The series in the right-hand side of equation (\ref{solm}) can be
calculated analytically for the initial {\em negative binomial
distribution\/} (the corresponding pure {\em negative binomial states\/}
were introduced independently in \cite{AharLer71,JoLa89,Mats})
\be
p_n^{(negbin)}(0)= \frac{\Gamma(\mu +n)\,\mu^{\mu}\,\overline{n}_0^{\,n}}
{\Gamma(\mu) n! \left(\overline{n}_0+\mu\right)^{\mu+n}}\,,
\quad \mu> 0,
\label{negbinin}
\ee
where $\overline{n}_0$ is the initial average number of photons.
The result is expressed in terms of the confluent hypergeometric function:
\br
&&p_n(\tau) = p_n(0) e^{-\tau}
\Phi\left(\mu+n;n+1;\frac{\overline{n}_0\,\tau}{\overline{n}_0 +\mu}\right)
\label{negbint1} \\
&&= p_n(0) \exp\left(- \frac{\mu\,\tau}{\overline{n}_0 +\mu}\right)
\Phi\left(1-\mu;n+1;
\frac{-\overline{n}_0\,\tau}{\overline{n}_0 +\mu}\right).
\label{negbint2}
\er
For integral values $\mu=1,2,\ldots$ formula (\ref{negbint2})
can be written in terms of the associated Laguerre polynomials:
\[
p_n(\tau) = \frac{\mu^{\mu}\,\overline{n}_0^{\,n}}
{ \left(\overline{n}_0\!+\!\mu\right)^{\mu+n}}
\exp\left(- \frac{\mu\,\tau}{\overline{n}_0 \!+\!\mu}\right)
L_{\mu-1}^{n}\left(
\frac{-\overline{n}_0\,\tau}{\overline{n}_0 \!+\!\mu}\right).
\]
The special case of $\mu=1$ corresponds to the initial {\em thermal\/}
distribution (coherent phase states in the case of pure quantum states)
\be
p_n(0)=\overline{n}_0^n/\left(1+\overline{n}_0\right)^{n+1}.
\label{therm-n}
\ee
In this case the photon number distribution preserves its form:
\be
p_n^{(th)}(\tau)=p_n(0)\exp\left[-\tau/\left(1+\overline{n}_0\right)\right],
\label{ptherm}
\ee
and the mean number of photons decreases with time exponentially, although
the rate of decrease diminishes with increase of the initial mean number:
\be
\overline{n}^{(th)}(\tau)=\overline{n}_0
\exp\left[-\tau/\left(1+\overline{n}_0\right)\right].
\label{ntherm}
\ee

A formal substitution $\mu\to -M$ with an integer $M$ transforms
the negative binomial distribution (\ref{negbinin})
to the {\em binomial\/} distribution
(pure {\em binomial states\/} were considered in \cite{AhaLer} and
rediscovered in \cite{binom,binLee})
\be
p_n^{(bin)}(0)=
\frac{M!\,\overline{n}_0^{\,n} \left(M-\overline{n}_0\right)^{M-n} }
{(M-n)! n! M^M}\,,
\quad \overline{n}_0 \le M.
\label{binin}
\ee
Then equation (\ref{negbint1}) is transformed to the formula
\[
p_n(\tau) = \left(\frac{\overline{n}_0}{M}\right)^{n}
\left(1-\frac{\overline{n}_0}{M}\right)^{M-n}
e^{-\tau}
L_{M-n}^{n}\left(
\frac{\overline{n}_0\,\tau}{\overline{n}_0 \!-\!M}\right).
\]

In the case of the initial {\em Poissonian\/} distribution
({\em coherent\/} pure states),
\be
p_n(0)=\frac{\overline{n}_0^n}{n!}\exp\left(-\overline{n}_0\right),
\label{Pois}
\ee
the series (\ref{solm}) is reduced to the modified Bessel function:
\be
p_n^{(Pois)}(\tau)=e^{-\overline{n}_0-\tau}
\left(\overline{n}_0/\tau\right)^{n/2}
I_n\left(2\sqrt{\overline{n}_0\tau}\right),
\label{pcoh}
\ee
\be
\overline{n}^{(Pois)}(\tau) = e^{-\overline{n}_0-\tau}\sum_{n=1}^{\infty}
n\left(\overline{n}_0/\tau\right)^{n/2}
I_n\left(2\sqrt{\overline{n}_0\tau}\right).
\label{ncoh}
\ee
The expressions (\ref{pcoh})--(\ref{ncoh}) can be simplified in the
asymptotical case $\overline{n}_0\tau \gg 1$, when the modified Bessel
functions can be replaced by exponentials. Actually, one needs an
additional condition $\tau \gg \overline{n}_0$ to ensure
that the simplified probabilities result in a convergent series whose value
does not exceed $1$. Thus for $\tau \gg \overline{n}_0 +\overline{n}_0^{-1}$
we can write
\be
p_n^{(Pois)}(\tau)\approx \frac{\left(\overline{n}_0/\tau\right)^{n/2}}
{\left(4\pi\sqrt{\overline{n}_0\tau}\right)^{1/2}}
\exp\left[-\left(\sqrt{\overline{n}_0} -\sqrt{\tau}\right)^2\right],
\label{pcohas}
\ee
\be
\overline{n}^{(Pois)}(\tau) \approx
\frac{\overline{n}_0^{1/4}}{\sqrt{4\pi}}\,\tau^{-3/4}
\exp\left[-\left(\sqrt{\overline{n}_0} -\sqrt{\tau}\right)^2\right].
\label{ncohass}
\ee

In figure 4 we compare $\overline{n}(\tau)/\overline{n}(0)$ for
Fock, thermal and coherent states from (\ref{avFock}),
(\ref{ntherm}) and (\ref{ncoh}), respectively, with the
corresponding result from amplitude damping model (\ref{ampd}).
While the amplitude damping model presents an exponential decay
independent of the field amplitude the exponential phase damping
does not, the field taking longer to relax in this situation. Only
for $k=1$ and Fock states the two processes coincide. On comparing
the curves for Fock, thermal and coherent states we observe that
the decay rate shows a strongly dependence on the field state
statistics. The more super-Poissonian is the field the longer will
it take to decay in the model based on the exponential phase
operators. It is interesting to note this dependence of the
relaxation process to the field statistics. In discussions of the
amplitude relaxation process ({\it e.g.} \cite{milburn3,zurek}),
this basis dependence is only attributed to coherence properties
of quantum fields. In this aspect the coherent state is said to be
selected from all the states as the more robust to decoherence
from amplitude damping model \cite{zurek}. We leave the discussion
of coherence properties of the exponential phase model to a future
publication, but we can anticipate from figure 4 that coherent
states are not the more robust to dissipation in this model.

\section{Conclusion}
Summarizing, in this paper we proposed modifications in the SD
photocount theory in order satisfy {\em all} the precepts, as
proposed by Srinivas and Davies for a consistent theory. Our
central assumption was the choice of the exponential phase
operators $E_-$ and $E_+$ as real `annihilation' and `creation'
operators in the photocounting process, instead of $a$ and
$a^\dag$. The introduction of those operators in the continuous
photocount theory, besides eliminating inconsistencies, leads to
new interesting results related to the counting statistics. A
remarkable result, which is responsible for all the physical
consistency of the model, is that in this new form an
infinitesimal photocount operation $J^{E}$ really takes out {\em one
photon\/} from the field, if the vacuum state is not
present. Consequently, the photocounting probability distribution
for a Fock field state is Poissonian, evidencing again the direct
correspondence of number of counted photons and number of photons
taken from the field.

We also have investigated the evolution of the field state when
photons are counted, but with no readout, leading to the
pre-selected state. The mean photon number change shows now (in
contrast to the exponential law obtained for the amplitude damping
model) a non-exponential law, which only depends on the condition
that photons are present in the field, independently of their mean
number.

An advantage of the proposed model is its mathematical consistence.
Since many of its predictions, especially those related to
multiphoton events, are significantly different from the predictions
of the SD theory, it can be verified experimentally.
One of the first questions which could be answered is: whether the
decrease of number of photons in the cavity due to their
continuous counting always obeys the exponential law (\ref{nmedioa})
(i.e., the
rate of change is proportional to the instantaneous mean number
of photons), or nonexponential dependences can be also observed
(for example, in the case of detectors with large dead times)?

We leave for a future work a detailed study of photocounting
processes in the presence of other incoherent (dissipative)
processes and discussion about coherence properties of the field
under the exponential phase damping model.
Also, an open question is how extremely nonlinear
(with respect to $a$ and $a^{\dagger}$) one-count operator
(\ref{J-E}) could be derived from some microscopical model,
starting from fundamental interactions which are linear with
respect to operators $a$ and $a^{\dagger}$ (as soon as the model considered
in our paper is pure phenomenological, as well as the SD model).

We would like to
emphasize that many preceding studies, such as
\cite{milburn1,milburn2,ueda,wiseman,Perin96,marcos}, adopted the
SD theory of photodetection. We believe that applications of our
consistent model of photodetection to these problems may, indeed,
bring new and important results for both the quantum measurement
theory and experiment.

\acknowledgments{This work was supported by FAPESP (S\~ao Paulo,
Brazil) under contracts  00/15084-5 and 01/00530-2}. {SSM also
acknowledges partial financial support by CNPq (Bras\'{\i}lia, DF).
VVD thanks CNPq for the full financial support.
We are also grateful to referees for a critical reading of the manuscript.}


\end{multicols}
\newpage

{\bf Figure captions}
\vspace{2cm}

{\bf Figure 1.} Conditional photocount probability distribution for the
initial number state with $m=5$ photons. Solid lines are for the present
model while dashed ones are for the original SD theory. Numbers above the
curves correspond to the $k$-event.
\vspace{2cm}

{\bf Figure 2.} Same as figure 1 for a coherent state with average
$|\alpha|^2=5$ photons.
\vspace{2cm}

{\bf Figure 3.} Same as figure 1 for a thermal state with average
${\ovl n}=5$ photons.
\vspace{2cm}

{\bf Figure 4.} Normalized mean number of photons
$n(t) \equiv \overline n(\gamma t)/\overline n(0)$
in the cavity under continuous measurement, for mean initial number of
photons $\overline n (0)= 1$, $5$ and $10$. Different line styles represent
different states for each mean initial photon number.
Three lower curves correspond, in the order from
bottom to top, to the Fock state (solid line), coherent state (dashed),
and thermal state (dotted), for $\overline n(0)=1$.
Three middle curves are related to the case of $\overline n(0)=5$
in the following order (from bottom to top): Fock state (dash-dotted line),
coherent state (dash-dot-dotted), thermal state (short dashed).
Three upper curves are related to the case of $\overline n(0)=10$ in the same
order as before: Fock state (short dotted line),
coherent state (short dash-dotted), thermal state (solid).
The lowest solid line (the Fock state with $\overline n(0)=1$)
coincides with the exponential decay for the amplitude damping model.


\begin{references}
%
\bibitem{Neumann}
von Neumann J 1932 {\em Mathematische Grundlagen der Quantenmechanik\/}
(Berlin: Springer)
\bibitem{mandel} Mandel L 1958 {\em Proc. Phys. Soc.} {\bf 72} 1037
\bibitem{mandel63} Mandel L 1963
{\em Progress in Optics\/} vol 2, ed. E Wolf (Amsterdam: North-Holland)
p 181
\bibitem{mandel2} Mandel L, Wolf E and Sudarshan E C G 1964
{\em Proc. Phys. Soc.} {\bf 84} 435
\bibitem{glauber} Glauber R J 1963 {\em Phys. Rev.} {\bf130} 2529
\bibitem{kk} Kelley P L and  Kleiner W H 1964 {\em Phys. Rev.} {\bf136} 316
\bibitem{mollow}  Mollow B R 1968 {\em Phys. Rev.} {\bf168} 1896
\bibitem{scully} Scully M O and Lamb W E Jr 1969 {\em Phys. Rev.}
{\bf179} 368
\bibitem{PL-rev} Pe\v{r}inov\'a V and Luk\v{s} A 2000
{\em Progress in Optics\/} {\bf 40} ed E Wolf (Amsterdam: Elsevier) p 115
\bibitem{davies} Srinivas M D and  Davies E B 1981 {\em Opt. Acta\/}
{\bf28} 981
\bibitem{milburn1} Milburn G J and  Walls D F 1984
{\em Phys. Rev.} A {\bf 30} 56
\bibitem{milburn2} Holmes C A, Milburn G J and  Walls D F 1989
{\em Phys. Rev.} A {\bf 39} 2493
\bibitem{ueda} Ueda M, Imoto N and Ogawa T 1990
{\em Phys. Rev.} A {\bf 41} 3891
\bibitem{wiseman} Wiseman H M and  Milburn G J 1993
{\em Phys. Rev.} A {\bf 47} 642
\bibitem{Perin96} Pe\v{r}inov\'a V, Luk\v{s} A  and K\v{r}epelka J 1996
{\em Phys. Rev.} A {\bf 54} 821
\bibitem{marcos} de Oliveira M C,  da Silva L F and Mizrahi S S 2002
{\em Phys. Rev.} A {\bf 65} 062314

\bibitem{Dakna}
Dakna M, Kn\"oll L and Welsch D-G 1998 {\em Europ. Phys. J.} D  {\bf 3} 295

\bibitem{LuH} Hong L 1999
{\em Phys. Lett.} A {\bf 264} 265

\bibitem{Wang00} Wang X G 2000
{\em Opt. Commun.}  {\bf 178} 365

\bibitem{Man-Q} Mandel L 1979 {\em Opt. Lett.} {\bf 4} 205

\bibitem{mizdod} Mizrahi S S and Dodonov V V 2002
{\em J. Phys. A: Math. Gen.} {\bf 35} 8847

\bibitem{Bal79} Baltes H P, Quattropani A and Schwendimann P  1979
{\em J. Phys. A: Math. Gen.} {\bf 12} L35

\bibitem{Lee} Lee C T 1993 {\em Phys. Rev.} A  {\bf 48} 2285

\bibitem{London} London F 1926 {\em Z. Phys.} {\bf 37} 915\\
London F 1927 {\em Z. Phys.} {\bf 40} 193

\bibitem{susg} Susskind L and Glogower J  1964 {\em Physics\/}  {\bf 1} 49

\bibitem{Niethis} Nieto M M 1993 {\em Phys. Scripta\/} {\bf T48} 5

\bibitem{CarNiet} Carruthers P and Nieto M 1968
{\em Rev. Mod. Phys.}  {\bf 40} 411
%
\bibitem{Paul}
Paul H 1974
{\em Fortschr. Phys.}  {\bf 22} 657
%
\bibitem{Imoto85} Imoto N, Haus H A and Yamamoto Y 1985
{\em Phys. Rev.} A  {\bf 32} 2287

\bibitem{BP86} Barnett S M and Pegg D T 1986
{\em J. Phys. A: Math. Gen.} {\bf 19} 3849

\bibitem{BeEng91}
Bergou J and Englert B-G  1991
{\em Ann. Phys.} (NY)  {\bf 209} 479
%
\bibitem{Loudon} Loudon R 1994 {\em The Quantum Theory of Light}
(Oxford: Clarendon).

\bibitem{LukPer}
Luk\v s A and Pe\v rinov\'a V 1994
{\em Quant. Opt.}  {\bf 6} 125
%
\bibitem{Ban95}
Ban M 1995 {\em Phys. Lett.} A {\bf 199} 275
\bibitem{Lynch95}
Lynch R 1995 
{\em Phys. Rep.}  {\bf 256} 367
\bibitem{Royer96}
Royer A 1996
{\em Phys. Rev.} A {\bf 53} 70

\bibitem{Ler} Lerner E C, Huang H W and Walters G E 1970 {\em J. Math. Phys.}
{\bf 11} 1679
\bibitem{Ifan} Ifantis E K 1972 {\em J. Math. Phys.}  {\bf 13} 568
\bibitem{ShapShep} Shapiro J H and Shepard S R 1991
{\em Phys. Rev.} A {\bf 43} 3795
\bibitem{ChaKa91} Chaturvedi S, Kapoor A K, Sandhya R, Srinivasan V
and Simon R 1991 {\em Phys. Rev.} A {\bf 43} 4555
\bibitem{Vourd92} Vourdas A 1992 {\em Phys. Rev.} A {\bf 45} 1943
\bibitem{Hall93} Hall M J W 1993  {\em J. Mod. Opt.} {\bf 40} 809
\bibitem{Sudar93} Sudarshan E C G 1993
{\em Int. J. Theor. Phys.}  {\bf 32} 1069
\bibitem{BrifBen94} Brif C and Ben-Aryeh Y 1994
{\em Phys. Rev.} A  {\bf 50} 3505
\bibitem{DoMi} Dodonov V V and Mizrahi S S 1995 {\em Ann. Phys.} (NY)
{\bf 237} 226
\bibitem{VBM96} Vourdas A, Brif C and Mann A 1996 {\em J. Phys. A: Math. Gen.}
{\bf 29} 5887
\bibitem{Wun01} W\"unsche A 2001 {\em J. Opt.} B {\bf 3} 206
\bibitem{rev-ncs} Dodonov V V 2002 {\em J. Opt.} B {\bf 4} R1

\bibitem{Moya99} Moya-Cessa H, Chavez-Cerda S and Vogel W 1999
{\em J. Mod. Opt.} {\bf 46} 1641

\bibitem{Lee97} Lee C T 1997
{\em Phys. Rev.} A  {\bf 55} 4449

\bibitem{SMW96} Scully M O, Meyer G M and Walther H 1996
{\em Phys. Rev. Lett.} {\bf 76} 4144

\bibitem{caves} Caves C M and Milburn G J 1987 {\em Phys. Rev.}
A {\bf 36} 5543

\bibitem{CoTa} Cohen-Tannoudji C, Dupont-Roc J and Grynberg G 1992
{\em Atom--Photon Interactions\/} (New York: Wiley)
\bibitem{Carm} Carmichael H 1993 {\em An Open Systems Approach to
Quantum Optics\/} (Berlin: Springer)
\bibitem{Calsa01} Calsamiglia J, Barnett S M, L\"utkenhaus N and
Suominen K-A 2001 {\em Phys. Rev.} A {\bf 64} 043814
\bibitem{AharLer71} Aharonov Y, Huang H W, Knight J M and Lerner E C 1971
{\em Lett. Nuovo Cim.} {\bf 2} 1317
\bibitem{JoLa89}  Joshi A and Lawande S V 1989 {\em Opt. Commun.} {\bf 70} 21
\bibitem{Mats} Matsuo K 1990 {\em Phys. Rev.} A   {\bf 41} 519
\bibitem{AhaLer} Aharonov Y, Lerner E C, Huang H W and Knight J M 1973
{\em J. Math. Phys.} {\bf 14} 746
\bibitem{binom}  Stoler D,  Saleh B E A and  Teich M C 1985 {\em Opt. Acta\/}
{\bf 32} 345
\bibitem{binLee} Lee C T 1985 {\em Phys. Rev.} A {\bf 31} 1213
\bibitem{milburn3} Walls D F and Milburn G J 1985 {\em Phys. Rev. A} {\bf 31} 2403
\bibitem{zurek} Zurek W H, Habib S and Paz J P 1993 {\em Phys. Rev. Lett.}
{\bf 70} 1187
\end{references}
\end{document}